\documentclass[aps,prl,reprint,groupedaddress,showpacs,nofootinbib]{revtex4-1}

\usepackage{graphicx}
\usepackage{amsmath}
\usepackage{amssymb}
\usepackage{aas_macros}
\usepackage{comment}

\newcommand{\msun}{\ensuremath{\mathrm{M}_\odot}}

\begin{document}
\title{Optical tomography of chemical elements synthesized in Type Ia supernovae}
\author{Ivo R. Seitenzahl}
\email{i.seitenzahl@adfa.edu.au}
\affiliation{School of Science, University of New South Wales, Australian Defence Force Academy, Canberra, ACT 2600, Australia}
\author{Parviz Ghavamian} 
\affiliation{Department of Physics, Astronomy and Geosciences, Towson University, Towson MD 21252, USA}
\author{J.~Martin Laming} 
\affiliation{Space Science Division Code 7684, Naval Research Laboratory, Washington DC 20375, USA}
\author{Fr\'ed\'eric P.~A.~Vogt}
\affiliation{European Southern Observatory, Av.~Alonso de C\'ordova 3107, 763 0355 Vitacura, Santiago, Chile}
\bibliographystyle{apsrev}

\date{\today}

\begin{abstract}
We report the discovery of optical emission from the non-radiative shocked ejecta of three young Type Ia supernova remnants (SNRs): SNR 0519-69.0, SNR 0509-67.5, and N103B. Deep integral field spectroscopic observations reveal broad and spatially resolved [Fe\,\textsc{xiv}] 5303\AA\ emission. The width of the broad line reveals, for the first time, the reverse shock speeds. For two of the remnants we can constrain the underlying supernova explosions with evolutionary models. SNR 0519-69.0 is well explained by a standard near-Chandrasekhar mass explosion, whereas for SNR 0509-67.5 our analysis suggests an energetic sub-Chandrasekhar mass explosion. With [S\,\textsc{xii}], [Fe\,\textsc{ix}], and [Fe\,\textsc{xv}] also detected, we can uniquely visualize different layers of the explosion. We refer to this new analysis technique as “supernova remnant tomography”.
\end{abstract}
\maketitle

Type Ia Supernovae (SNe Ia) are the thermonuclear explosions of white dwarf stars. In spite of their importance as distance indicators in Cosmology \cite{riess1998a,perlmutter1999a} and their major contribution to nucleosynthesis \cite{seitenzahl2017a}, no consensus has been reached on their explosion mechanism(s) and progenitor system(s) \cite{hillebrandt2013a}. Even for the well-studied, nearby SN 2011fe, a comparison of the observations and synthetic spectral time series of the two leading explosion models has failed to produce a clear winner: the “single degenerate” delayed-detonation model of a ${\sim}1.4\,\msun$ WD \cite{seitenzahl2013a} and the “double-degenerate” merger with a ${\sim}1.1\,\msun$ \cite{pakmor2012a} primary WD explain the observations nearly equally well \cite{roepke2012a}. \\
\indent An alternative approach to solving the SN Ia progenitor problem is via multi-wavelength observations of supernova remnants (SNRs). Following the thermonuclear incineration of a white dwarf, the freshly synthesized heavy elements are ejected at high velocity. The supersonic expansion drives a forward shock into the surrounding interstellar medium and a reverse shock back into the remains of the supernova explosion, eventually heating the ejecta to X-ray emitting temperatures \cite{reynolds2008a}. The most important parameters governing the evolution of SNRs are the chemical composition, kinetic energy and mass of the ejecta, as well as the ambient medium density \cite{truelove1999a}, all of which are closely linked to the explosion mechanism. As the supernova ejecta progressively ionize behind the reverse shock, zones of higher and higher atomic ionization are produced in succession behind this shock. Optical forbidden line emission from low-lying atomic transitions of these highly-ionized atoms is expected. Many of these lines were first seen in the solar corona and are hence referred to as “coronal” lines. \\
\indent The coronal [Fe\,\textsc{xiv}] magnetic dipole transition 3s$^{2}$3p$^{2}$ (P$_{1/2} - $ P$_{3/2}$) produces a green emission line at 5302.8\AA\, with an emissivity that peaks in ionization equilibrium at temperatures near $2\times10^6$\,K \cite{bryans2006a} and is produced over the range $7.0 < \log(\mathrm{T}) < 7.5$ in the shock models presented below. Earlier detections of [Fe\,\textsc{xiv}] in SNRs were from “radiative” cloud shocks in ISM material (${\sim}300 - 500\,\mathrm{km}\,\mathrm{s}^{-1}$, where the postshock gas undergoes thermal instability and the shock dynamics are strongly affected by radiative cooling), such as those detected in Puppis A \cite{lucke1979a,teske1987a}, N49 \cite{dopita2016a}, and 1E\,0102.2-7219 \cite{vogt2017a}, following model predictions \cite{allen2008a}. In these cases, the sensitivity of the detectors has been the limiting factor in detecting optical [Fe\,\textsc{xiv}] from the much faster non-radiative shocks ($>2000\,\mathrm{km}\,\mathrm{s}^{-1}$, no thermal instability) in both the swept up interstellar gas and reverse shocked ejecta.
As we show in this paper, the superior sensitivity of the Multi Unit Spectroscopic Explorer (MUSE) Integral field spectrograph on the European Southern Observatory (ESO) Very Large Telescope (VLT) and the larger light gathering area of its 8.2\,m mirror have now enabled the detection of faint optical coronal line emission in non-radiative shocks. Using public MUSE data from the ESO archive, we have discovered [Fe\,\textsc{xiv}] 5303\AA\ emission from the reverse shocks of the three youngest Type Ia supernova remnants in the Large Magellanic Cloud (LMC) \cite{hughes1995a}: SNR 0519-69.0, SNR 0509-67.5, and N103B (SNR 0509-68.7). For further details on the observations, data reduction and processing see the Supplemental Information.

\begin{figure*}[th!]
  \includegraphics[width=\textwidth]{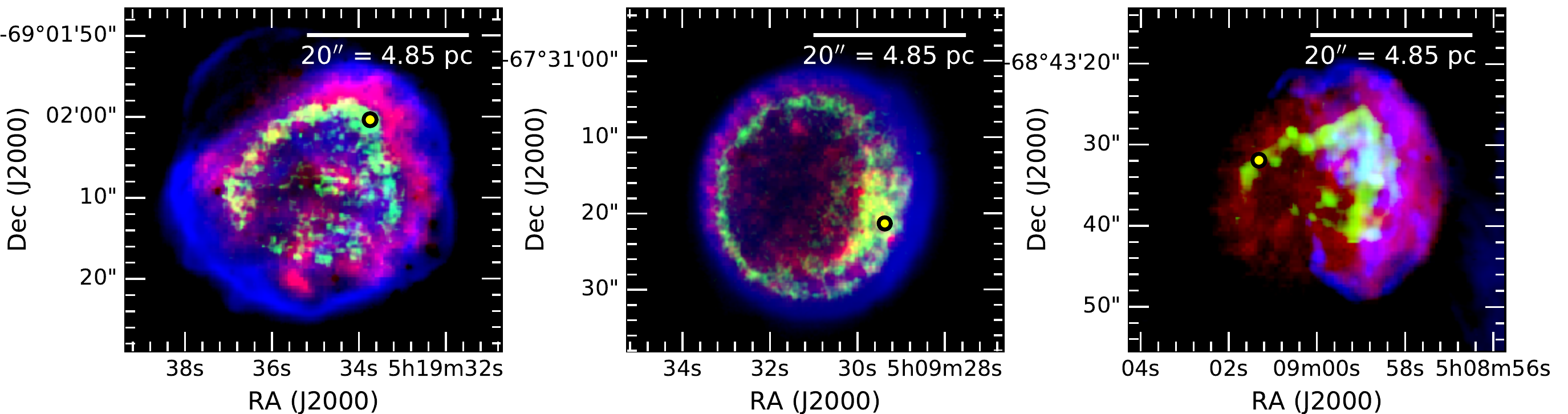}
  \caption{RGB images of 0519-69.0 (A), 0509-67.5 (B) and N103B (C) showing in red X-rays from Chandra ACIS, in blue H$\alpha$ (MUSE), and in green [Fe\,\textsc{xiv}] (MUSE). The regions from which the spectra were extracted are indicated by the yellow dots.}
  \label{fig:f1}
\end{figure*}

\begin{figure}[th!]
  \includegraphics[width=9.5cm,trim={0.7cm 0 0 0}, clip]{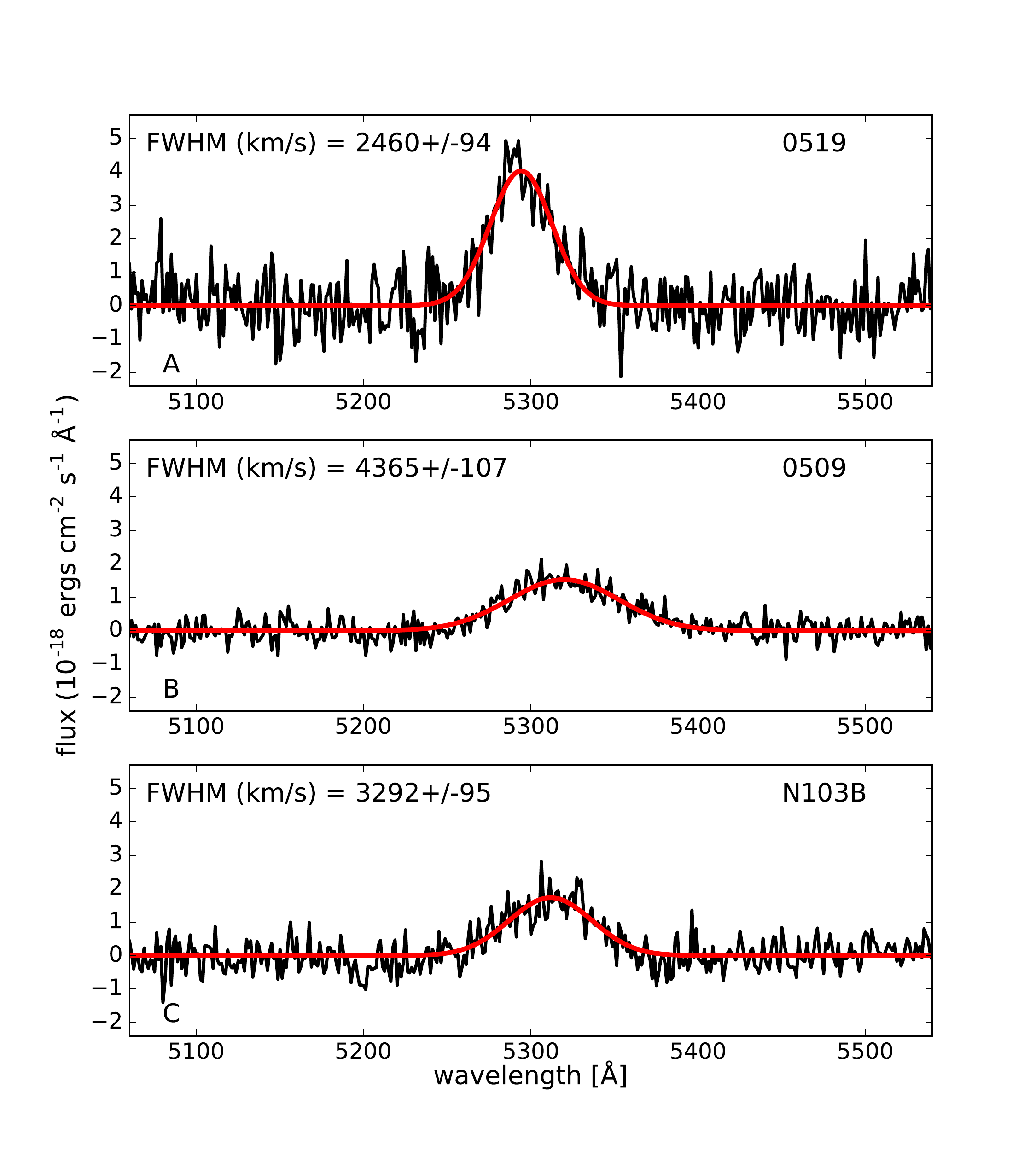}
  \caption{[Fe\,\textsc{xiv}] 5303 line profiles for 0519-69.0 (A), 0509-67.5 (B), and N103B (C) extracted from the regions indicated (yellow dots with black edges) in Figure 1. The apertures are circular with 0.8 arcsecond radii, corresponding to 1.96 square arcsecond areas (49 MUSE spaxels). Shown in red are best-fitting Gaussians to the data, determined by a least-squares minimization.}
  \label{fig:f2}
\end{figure}

\begin{figure*}[th!]
  \includegraphics[width=\textwidth]{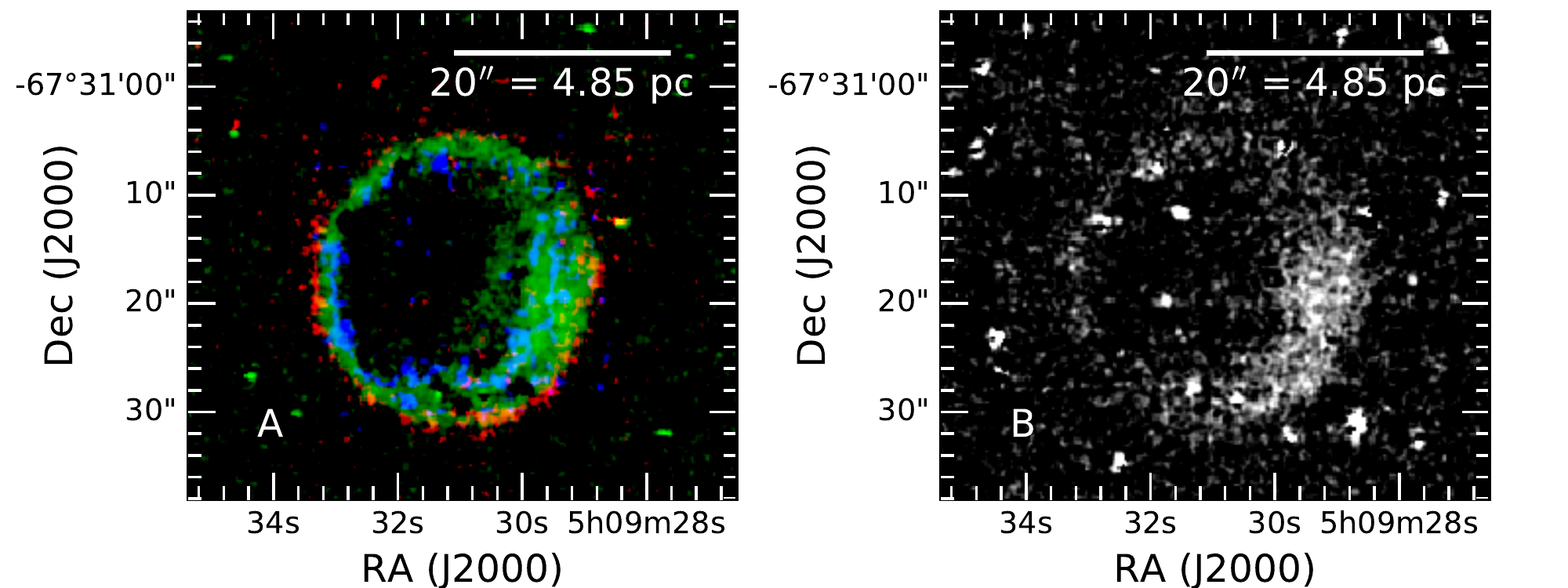}
  \caption{(A) Left panel is RGB image of 0509-67.5 showing in red [S\,\textsc{xii}], in blue [Fe\,\textsc{ix}], and in green [Fe\,\textsc{xiv}]. (B) Right panel is image of 0509-67.5 in [Fe\,\textsc{xv}].}
  \label{fig:f3}
\end{figure*}

To our knowledge this is the first detection of optical emission from the non-radiatively shocked ejecta of any Type Ia supernova remnant. As expected, we find the peak of the [Fe\,\textsc{xiv}] emission, which appears as a narrow band in the interiors of the SNRs, immediately interior to the peak of the Fe K X-ray emission detected by the Chandra X-ray Observatory (see Fig.~1), since the reverse shock is propagating inwards in a Lagrangian sense. The detection of optical coronal line emission from pure non-radiative ejecta shocks of Type Ia SNRs opens a long sought window into the kinematic study of young Type Ia SNRs. 
In the case of SNR 0519-69.0 (hereafter 0519) and SNR 0509-67.5 (hereafter 0509), the [Fe\,\textsc{xiv}] emission appears as a nearly circular shell (see Fig.~1(A) and (B)). For N103B (Fig.~1(C)), the signal is contaminated by residuals from superimposed bright stars and the [Fe\,\textsc{xiv}] behind the reverse shock appears much more asymmetric. This is likely a consequence of the strong interaction of this SNR with high-density material on its western side \cite{williams2014a}. The observed morphology of three nearly concentric shells of Balmer-line emission from the forward shock (blue) on the outside, with X-ray emission (red) from the hot, reverse shocked ejecta just inside the Balmer filaments, and coronal [Fe\,\textsc{xiv}] emission (green) inside of the X-ray emitting ejecta, is a beautiful confirmation of the extant theory of SNR evolution. To probe the kinematics of the iron-rich ejecta in each SNR, we have extracted [Fe\,\textsc{xiv}] line profiles (see Fig.~2) from selected regions (indicated in Fig.~1) of the three SNRs. Fitting single Gaussians and a linear continuum to the line profiles, we obtain velocity widths of $2460\pm100\,\mathrm{km}\,\mathrm{s}^{-1}$ for 0519, $4370\pm100\,\mathrm{km}\,\mathrm{s}^{-1}$ for 0509, and $3290\pm100\,\mathrm{km}\,\mathrm{s}^{-1}$ for N103B.

The near spherical symmetry of 0519 and 0509 allows us to model them in 1D (whereas the strongly asymmetric morphology of the [Fe\,\textsc{xiv}] in N103B does not), so in the remainder of this report we focus on these two SNRs for a quantitative analysis. While the approximate location of the reverse shock can be inferred from X-ray observations of the shocked ejecta \cite{kosenko2010a}, the resolved line width of the [Fe\,\textsc{xiv}] emission presented here allows us for the first time to directly determine the reverse shock speed \textendash\ a new observational constraint. The radius of the peak of the [Fe\,\textsc{xiv}] emission, modeled as a spherical shell, is $2.86\pm0.10\,$pc for 0509 and $2.36\pm0.18\,$pc for 0519, respectively. To provide estimates of the total line fluxes we integrated the broad [Fe\,\textsc{xiv}] line over the full extent of the emission in each SNR and fit a single Gaussian to each line profile after subtracting a linear continuum. Corrected for extinction and reddening by dust, we obtain estimates of total line fluxes of $1.1\times10^{-14}\,\mathrm{erg}\,\mathrm{cm}^{-2}\,\mathrm{s}^{-1}$ for 0519 and $0.9\times 10^{-14}\,\mathrm{erg}\,\mathrm{cm}^{-2}\mathrm{s}^{-1}$ for 0509. 

A valuable constraint on the interpretation of our [Fe\,\textsc{xiv}] measurements is found in the time evolution of observed light echoes \textendash\ the reflections of supernova light by interstellar dust sheets. Modeling of the light echoes [19] allowed for an explosion model and SNR evolution independent determination of the SNR ages. These models placed 0519 at $600\pm200$\,yr and 0509 at $400\pm120$\,yr \cite{rest2005a}. Further, since these two SNRs are located in the LMC, their distances are reliably known to be 50\,kpc, with an uncertainty of only 2 per cent \cite{pietrzynski2013a}. This allows us to accurately relate angular size to physical size. The forward shock position and velocity can be inferred from the broad Balmer-line emission \cite{hovey2015a,hovey2018a}. With reliable observational constraints on the age, forward shock position and velocity, as well as reverse shock position and velocity, we are now in a position to limit explosion model parameters commensurate with the observational constraints.

\begin{table}[h]
\scriptsize
\caption{\label{table1}Explosion energies $E_{51}$ ($10^{51}\,\mathrm{erg}$), ejecta masses $M_{ej}$ (solar masses), ISM densities $n_{ISM}$ ($\mathrm{amu}\,\mathrm{cm}^{-3}$), assumed age, modeled forward and reverse shock velocities and radii, and for 0519 ejecta ionization age and electron temperature, compared with observational values from literature references. Modeled and observed radii, width, [Fe\,\textsc{xiv}] 5303\AA\ flux, Fe mass associated with [Fe\,\textsc{xiv}] emission (includes [Fe\,\textsc{x}] also for 0509), and estimate of total SNR Fe mass. }
     \begin{ruledtabular}
     \begin{tabular}{cc c c c}
     SNR & \multicolumn{2}{c}{0519: $E_{51}=1$;}  & \multicolumn{2}{c}{0509: $E_{51}=1.5$;}  \\
         & \multicolumn{2}{c}{$M_{ej}=1.4$; $n_{ISM}=1.5$} & \multicolumn{2}{c}{$M_{ej}=1.0$; $n_{ISM}=0.4$}  \\
     \hline
         & observation & model & observation & model \\ 
         \hline
      Age (yr) & $600\pm200$ \cite{rest2005a} & 750 & $400\pm120$  \cite{rest2005a} & 310 \\
               &              &     & $310\pm35$ \cite{hovey2015a}  &   \\
      $v_f (\mathrm{km}\,\mathrm{s}^{-1}$) & $2770\pm500$ \cite{kosenko2010a} & 2516 & $6500\pm200$ \cite{hovey2015a} & 6539 \\
                      & 2650 \cite{hovey2018a} &  &  & \\
     $R_f$ (pc) & $4.0\pm0.3$ \cite{kosenko2010a} & 4.07 & 3.636 \cite{hovey2015a} & 3.64 \\
     $v_r (\mathrm{km}\,\mathrm{s}^{-1}$) &  & 1887 &  & 4766 \\
     $R_r$ (pc) &  & 2.16 &  & 2.74 \\
     $v_{exp} (\mathrm{km}\,\mathrm{s}^{-1}$) &  & 4057 &  & 5170 \\
     $n_{e}t$  & $3.8\pm0.3$ \cite{kosenko2010a} & 3.7 & 0.85 \textendash\ 3.4 \cite{badenes2008a} & 0.315 \\
      $(10^{10}\,\mathrm{cm}^{-3}\,\mathrm{s})$ & & & 1.4 \textendash\ 1.6 \cite{warren2004a} & \\
      $T_e$ (K) & 3.2e7 \cite{kosenko2010a} & 5.1e7 & $3.6\pm0.6$e7 \cite{badenes2008a} & 1.97e7 \\
      & & &4.6 \textendash\ 5.8e7 \cite{warren2004a} & \\
      $R_{Fe\textsc{xiv}}$  & 2.18 \textendash\ 2.55 & 2.8 \textendash\ 2.9 & 2.76 \textendash\ 2.96 & 2.81 \textendash\ 2.85 \\
      (pc) & & & & \\
      $W_{Fe\textsc{xiv}}$  & $2460\pm94$ & 3600 & $4365\pm107$ & 5117 \\
      ($\mathrm{km}\,\mathrm{s}^{-1}$)& & & & \\
      $F_{Fe\textsc{xiv}}$  & 1.1e-14 &  & 0.9e-14 &  \\
      $(\mathrm{erg}\,\mathrm{cm}^{-2}\,\mathrm{s}^{-1})$ & & & & \\
      $M_{Fe\textsc{xiv}}$  &  & 0.03 &  & 0.015 \\
      $(M_{\odot})$ & & & & \\
      $M_{Fetot}$  &  & 0.38 &  & 0.515 \\
      $(M_{\odot})$ & & & & \\
     \end{tabular}
     \end{ruledtabular}
     \end{table}

Table 1 gives a summary of our SNR models based on \cite{truelove1999a} for 0519 and 0509, designed to match forward and reverse shock velocities ($v_f$ and $v_r$) and radii ($R_f$ and $R_r$) at the current epoch. The ejecta mass is $M_{ej}$, $E_{51}$ is the explosion kinetic energy in $10^{51}\,\mathrm{erg}$, and $n_{ISM}$ is the ambient density in $\mathrm{amu}\,\mathrm{cm}^{-3}$. For 0519 we take $M_{ej} = 1.4\,\msun$, E$_{51} = 1$, and a chemical composition 33\% O, 12\% Si and 55\% Fe mass, to match the results of X-ray analysis \cite{kosenko2010a}. With $n_{ISM} = 1.5\,\mathrm{cm}^{-3}$ taken to match the forward shock, we also get good agreement for the ejecta ionization age and electron temperature. Currently the reverse shock in 0519 has passed through approximately 95\% of the ejecta (mass coordinate ~0.05), and the [Fe\,\textsc{xiv}]-emitting plasma is near mass coordinate 0.2, expanding with $v_{exp} = 1887\,\mathrm{km}\,\mathrm{s}^{-1}$. For further details on our SNR hydrodynamical evolution model and ionization structure calculations see the Supplemental Information.

In the case of 0509, while adopting an explosion energy $E_{51} = 1.5$ and $n_{ISM} = 0.4\,\mathrm{cm}^{-3}$ allows the forward shock radius and velocity to be matched as well as the emission measure of shocked ISM, a similar ejecta mass and composition to 0519 do not allow the Fe to ionize as far as $\mathrm{Fe}^{13+}$. However, a smaller ejecta mass $M_{ej}$ = 1\,\msun\ allows the reverse shock to reach the ejecta core-envelope boundary, where the maximum ionization age occurs, earlier in the SNR evolution. This produces sufficient $\mathrm{Fe}^{13+}$ and $\mathrm{Fe}^{14+}$ here to generate brighter [Fe\,\textsc{xiv}] 5303\AA\ than [Fe\,\textsc{xi}] 7892\AA\ or [Fe\,\textsc{x}] 6376\AA, neither of which are unambiguously detected. We note that this high explosion energy is realistic and can be readily obtained from detonation of a 1\,\msun\ white dwarf with a 0.85\,\msun\ core consisting of 60\% carbon and 40\% oxygen (by mass) surrounded by a 0.15\,\msun\ shell of helium. Burning 0.5\,\msun\ of the core to iron-group elements (using the binding energy of $^{56}$Ni) and the remainder of the star to intermediate mass elements (using the binding energy of $^{28}$Si) gives a kinetic energy of $1.5\times10^{51}\,\mathrm{erg}$, after accounting for the gravitational binding energy $E_{g} = -4.6\times 10^{50}\,\mathrm{erg}$ and the internal energy $E_{int} = 2.9\times 10^{50}\,\mathrm{erg}$.

For the 1\,\msun\ ejecta model, the reverse shock in 0509 has passed through approximately 74\% of the ejecta (mass coordinate 0.26) at the present time, and the [Fe\,\textsc{xiv}] emission originates from mass coordinates ${\sim}0.5 - 0.7$, expanding with $v_{exp} = 4766\,\mathrm{km}\,\mathrm{s}^{-1}$. Table 1 gives a summary of parameters connected with the [Fe\,\textsc{xiv}] emission for both remnants. There is good agreement between predicted and observed radii, with the observations giving a wider range of values. Presumably this arises partly from simple projection effects and partly from deviations of the SNR geometry from spherical symmetry. The line widths, however, are over-predicted by about 10 \textendash\ 20\%. The theoretical prediction is directly connected to the speed of the reverse shock and is possibly affected by the parametrization of the ejecta density profile by a uniform density core, or by clumping of the ejecta, which would slow down the reverse shock. 

In Table 1, the de-reddened fluxes in [Fe\,\textsc{xiv}] are given for the two remnants, with an estimate of the Fe mass in all charge states associated with the [Fe\,\textsc{xiv}] emission, coming from our ionization balance calculations. The final row of Table 1 gives an estimate of the total Fe in each remnant. To the Fe associated with [Fe\,\textsc{xiv}], we add the mass of currently unshocked ejecta ($0.18 \times 1.4 = 0.25\,\msun\ $ for 0519, $0.5 \times 1.0 = 0.5\,\msun\ $ for 0509), assumed all Fe, and for 0519 we add estimates of the shocked Fe mass seen in X-rays [18]. For further details on the Fe mass estimate from the observed line flux see the Supplemental Information.

The characteristic velocity, distance, and time in our models depend on $(E_{51}/M_{ej})^{1/2}$, $(M_{ej}/n_{ISM})^{1/3}$, and $M_{ej}^{5/6}E_{51}^{-1/2}n_{ISM}^{-1/3}$, respectively, so in Table 1 only $n_{e}t$ and $T_e$ change if $E_{51}$, $M_{ej}$, and $n_{ISM}$  vary by the same factor. A factor of ${\sim}4$ increase in $n_{e}t$ is required to improve the agreement between predicted and measured $n_{e}t$ for 0509, which conflicts with established Type Ia SN theory. If we solely increase $M_{ej}$ and the age for 0509, $n_{e}t$ and $T_{e}$ increase somewhat, but simultaneously $v_f$ decreases and $R_f$ increases, worsening the prediction of the forward shock trajectory. A modest increase in $M_{ej}$ by ${\sim}0.2 - 0.4\,\msun\ $ is allowable but would require the Fe to be embedded in He-rich ejecta to achieve the necessary degree of ionization. Such a near Chandrasekhar-mass scenario with unburned helium in the ejecta seems unlikely, but we cannot firmly rule out a near Chandrasekhar-mass explosion as for example in \cite{badenes2008a}. The larger mass makes the reverse shock slower, brings the [Fe\,\textsc{xiv}] width into better agreement with observations, and increases our estimate for the total Fe mass because the slower reverse shock has not propagated as far through the ejecta. However, the most satisfactory explanation for the $n_{e}t$ values is that the strong Si, S, Ar, and Ca emission seen in X-rays \cite{warren2004a} arises from ejecta clumps, with densities locally enhanced by a factor of ${\sim}4$. This gives a predicted $n_{e}t$ of order $10^{10}\,\mathrm{cm}^{-3}\,\mathrm{s}$. Using an electron density of $4\,\mathrm{cm}^{-3}$ to interpret the emission measures given in \cite{warren2004a} then yields masses of clumped ejecta of 0.068, 0.035, 0.007, and 0.003\,\msun\ for Si, S, Ar, and Ca, respectively, implying that a total of about 0.11\,\msun\ out of a total shocked ejecta mass of about 0.74\,\msun\ is clumped by a factor of about 4. Approximately 0.2\,\msun\  of the shocked ejecta mass is then visible in [Fe\,\textsc{xiv}] and [Fe\,\textsc{xv}]. Therefore, we favor the low mass – high explosion energy scenario.

A remaining question is why 0509 exhibits clumpy ejecta while 0519 apparently does not. Aside from being more than twice as old as 0509, 0519 is in a significantly more advanced evolutionary state due to its higher ambient density. Presumably, all ejecta clumps in 0519 have been destroyed by instabilities following reverse shock passage \cite{pittard2016a,orlando2010a}, whereas this has not yet occurred in 0509. Kelvin-Helmholtz and Richtmyer-Meshkov instabilities typically destroy clumps on a timescale of a few clump shock crossing times. Clumping of Fe in 0509 would remove the need for a He-dominated composition in the 1.4\,\msun\ model for explaining the Fe ionization, but poses problems in that clumping of SN ejecta is usually assumed to occur as a result of the inflation of Fe-Co-Ni bubbles by radioactivity. Fe should therefore be under-dense, though \cite{wang2001a} interpret Fe knots as being due to $^{54}$Fe.

In addition to the ubiquitous [Fe\,\textsc{xiv}] emission, we also find three additional broad lines in 0509, which we identify as coronal [S\,\textsc{xii}] 7613.1\AA, [Fe\,\textsc{ix}] 8236.8\AA,  (Fig.~3A) and [Fe\,\textsc{xv}] 7062.1\AA\ (Fig.~3B). We also detect [Fe\,\textsc{xv}] 7062.1\AA\ in N103B. The presence of these further coronal lines in addition to [Fe\,\textsc{xiv}] opens the door to a new field of study: supernova remnant tomography, the study of spatially resolved, optical coronal line emission from non-radiative reverse shocks in Type Ia supernova ejecta. The energetics of SNRs means that most of the emission from shocked ejecta is radiated at X-ray frequencies, observed with relatively poor spectral and spatial resolution due to technical limitations on the available instrumentation. Study of the optical coronal line profiles allows for the measurement of Doppler shifts and broadening. Furthermore, since the emission arises from much closer to the reverse shock than the X-ray emission, it is more sensitive to shock and pre-shock parameters. In contrast, the X-ray observations probe only the clumped ejecta, providing a less accurate picture of the spatial distribution of explosion products than the optical [Fe\,\textsc{xiv}] emission.

In the cases discussed here, the best match to the forward and reverse shocks pushes the SNR age to one end or the other of the uncertainty range coming from the light echoes and constrains the ejecta masses to around 1.4\,\msun\ for 0519 and likely to significantly below the Chandrasekhar mass for 0509 (${\sim}1.0\,\msun$). In the absence of such information, the SNR age is much less constrained, with corresponding greater uncertainties in ejecta mass and explosion energy. Our dynamical models give a good match to the spectral properties of 0519 and 0509, with some clumping of the ejecta required for the latter SNR. 
Last, we note that the observed light echo spectra enabled \cite{rest2008a} to assign the supernova that gave rise to 0509 to the spectroscopic sub-class of 1991T-like SNe Ia. Taking our explosion mass constraint at face value, this indicates that 1991T-like SNe Ia originate from detonations of sub-Chandrasekhar mass white dwarfs. 

\section{Acknowledgements}
  This research has made use of the following PYTHON packages: MATPLOTLIB \cite{hunter2007a}, ASTROPY \cite{astropy2013a,astropy2018a}, a community-developed core PYTHON package for Astronomy APLPY \cite{robitaille2012a}, an open-source plotting package for PYTHON, ASTROQUERY \cite{ginsburg2017a}, a package hosted at https://astroquery.readthedocs.io which provides a set of tools for querying astronomical web forms and databases, STATSMODEL \cite{seabold2010a} and BRUTIFUS \cite{vogt2019a}, a PYTHON module to process data cubes from integral field spectrographs. This research has also made use of the ALADIN \cite{bonnarel2000a} interactive sky atlas, of SAOIMAGE DS9 \cite{joye2003a} developed by Smithsonian Astrophysical Observatory, of NASA's Astrophysics Data System, and of NASA/IPAC Extragalactic Database (NED), which is operated by the Jet Propulsion Laboratory, California Institute of Technology, under contract with the National Aeronautics and Space Administration. IRS was supported by the Australian Research Council through grant number FT160100028. PG acknowledges support from the rector funded visiting fellowship scheme at the University of New South Wales in Canberra. JML was supported by the NASA ADAP Program Grant NNH16AC24I and by Basic Research Funds of the Chief of Naval Research. FPAV acknowledges an ESO fellowship.

\section{Author contributions:} Conceptualisation: IRS; PG; Formal Analysis: IRS, JML; Investigation: IRS, PG, JML; Methodology: IRS, PG, JML; Project administration: IRS; Resources: FPAV; Software: IRS, JML, FPAV; Visualization: IRS, JML, FPAV; Writing – original draft: IRS, PG, JML, FPAV; Writing – review \& editing: IRS, PG, JML, FPAV.

\bibliography{bibliography}{}
\clearpage

\onecolumngrid
\begin{center}
    \Large{\textbf{Supplemental material}}\\
\end{center}
\vspace{\columnsep}
\twocolumngrid

\section{MUSE observations and data reduction}

SNR 0519-69.0 was observed with the Multi-Unit Spectroscopic Explorer (MUSE) on UT4 at the Very Large Telescope (VLT), under P.Id.~096.D-0352[A] (P.I: Leibundgut), on 17.01.2016 and 18.01.2016, for a total $6\times900\,\mathrm{s}$ on-source. SNR 0509-67.5 was observed with MUSE, under P.Id. 0101.D-0151[A] (P.I.: Morlino), on 21.11.2017, 15.12.2017, and 22.01.2018, for a total of $16 \times 701$\,s on-source. SNR N103B was observed with MUSE, under P.Id. 096.D-0352[A] (P.I: Leibundgut), on 12.12.2016 and 17.12.2016, for a total of $16 \times 900$\,s on-source. We downloaded the 6, 16, and 16 raw MUSE frames for SNR 0519-69.0, SNR 0509-67.5, and SNR N103B (respectively) from the ESO Science Archive Facility, together with the associated raw calibration selected via calselector. Each raw frame was reduced individually using the MUSE pipeline 2.4.2 \cite{weilbacher2015a} via its workflow in Reflex v2.9.1 \cite{freudling2013a}. The image quality was measured manually from stars in the field-of-view, in all the individual datacubes, at 7000\AA. For SNR 0519-69.0 and SNR N103B, all the individual observations have an image quality in the range 0.8$^{\prime\prime}$ \textendash\ 0.9$^{\prime\prime}$ and 0.5$^{\prime\prime}$ \textendash\ 0.8$^{\prime\prime}$  (respectively, so that we used them all to assemble the final combined cubes, which have an image quality of 0.8$^{\prime\prime}$ and 0.6$^{\prime\prime}$ for a total of 5,400\,s and 14,400\,s on-source, respectively. For SNR 0509-67.5, the first 8 exposures have an image quality in the range 1.0$^{\prime\prime}$ \textendash\ 1.5$^{\prime\prime}$, whereas the last 8 exposures, acquired in January 2018, have an image quality in the range 0.7$^{\prime\prime}$ \textendash\ 0.9$^{\prime\prime}$. To facilitate the removal of the stellar continuum in the field (see below) and maximize our ability to spatially resolve the shell structure of the remnant, we only use the 8 sharpest MUSE exposures to assemble the combined datacube, with a final image quality of 0.8$^{\prime\prime}$.

The lack of dedicated sky observations for all 3 targets led us to skip the sky subtraction step in the data reduction cascade, given the crowding of the fields and the underlying photo-ionized gaseous emission from the LMC. Instead, we use brutifus \textemdash\ a Python package to process MUSE datacubes (https://fpavogt.github.io/brutifus/) \textemdash\ to subtract the sky emission extracted from a handful of regions selected by hand in each cube (see Supplemental Figs.~4-6). These regions were chosen a) to avoid any bright star in the white-light image of the cube, b) to be located away from the extend of the SNRs, and c) to avoid the brightest region of nebular emission from the LMC ISM. 

\begin{figure}[h!]
  \includegraphics[width=0.8\columnwidth,trim={0 4.2cm 0 6.8cm}, clip]{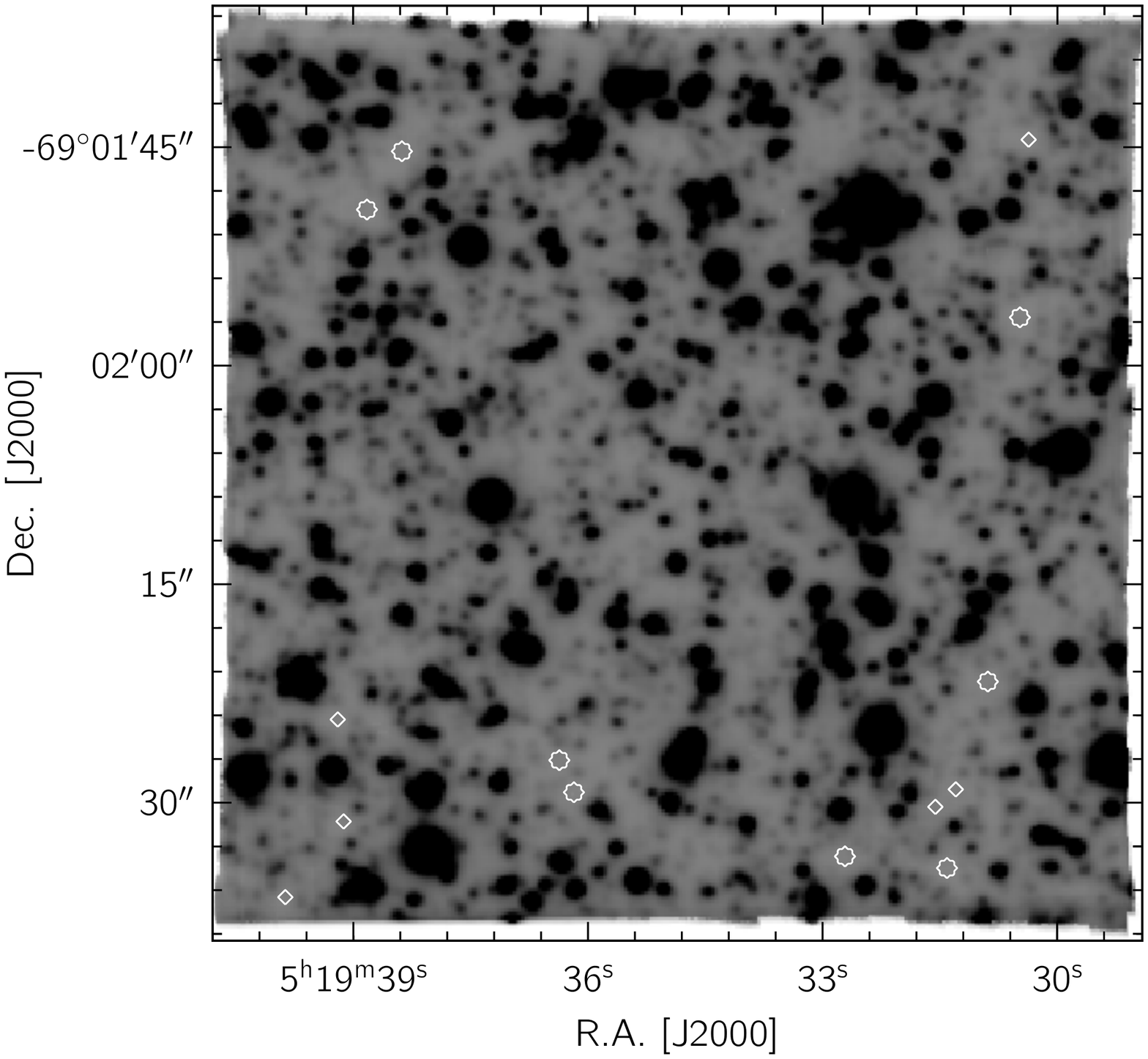}
    \vspace{-0.5cm}
  \caption{Regions used for the sky-subtraction procedure for SNR 0519-69.0.}
  \label{sfig:f1}
\end{figure}
\begin{figure}[h!]
  \includegraphics[width=0.8\columnwidth,trim={0 4.2cm 0 6.8cm}, clip]{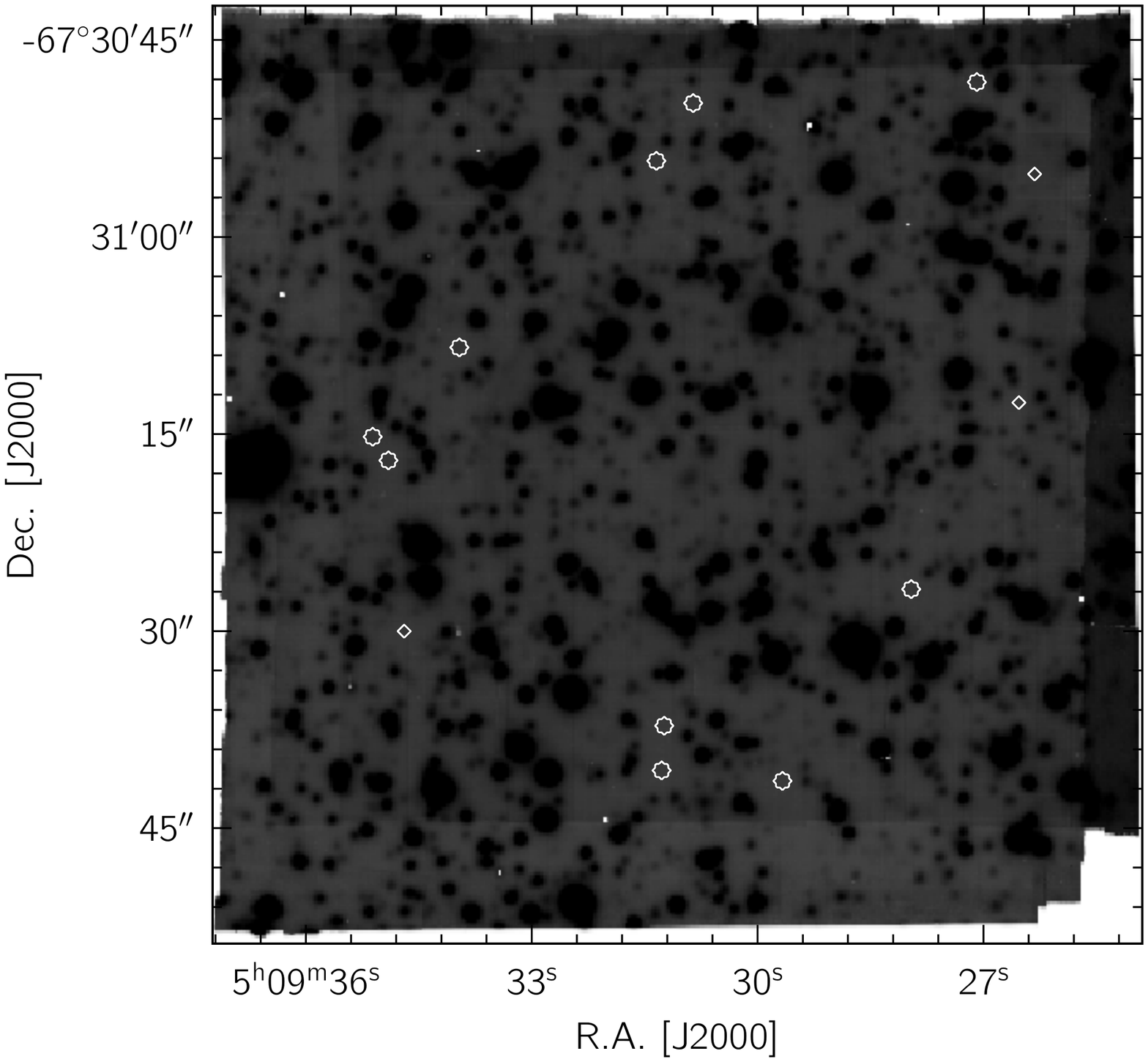}
    \vspace{-0.5cm}
  \caption{Regions used for the sky-subtraction procedure for SNR 0509-67.5.}
  \label{sfig:f2}
\end{figure}
\begin{figure}[h!]
  \includegraphics[width=0.8\columnwidth,trim={0 4.2cm 0 6.8cm}, clip]{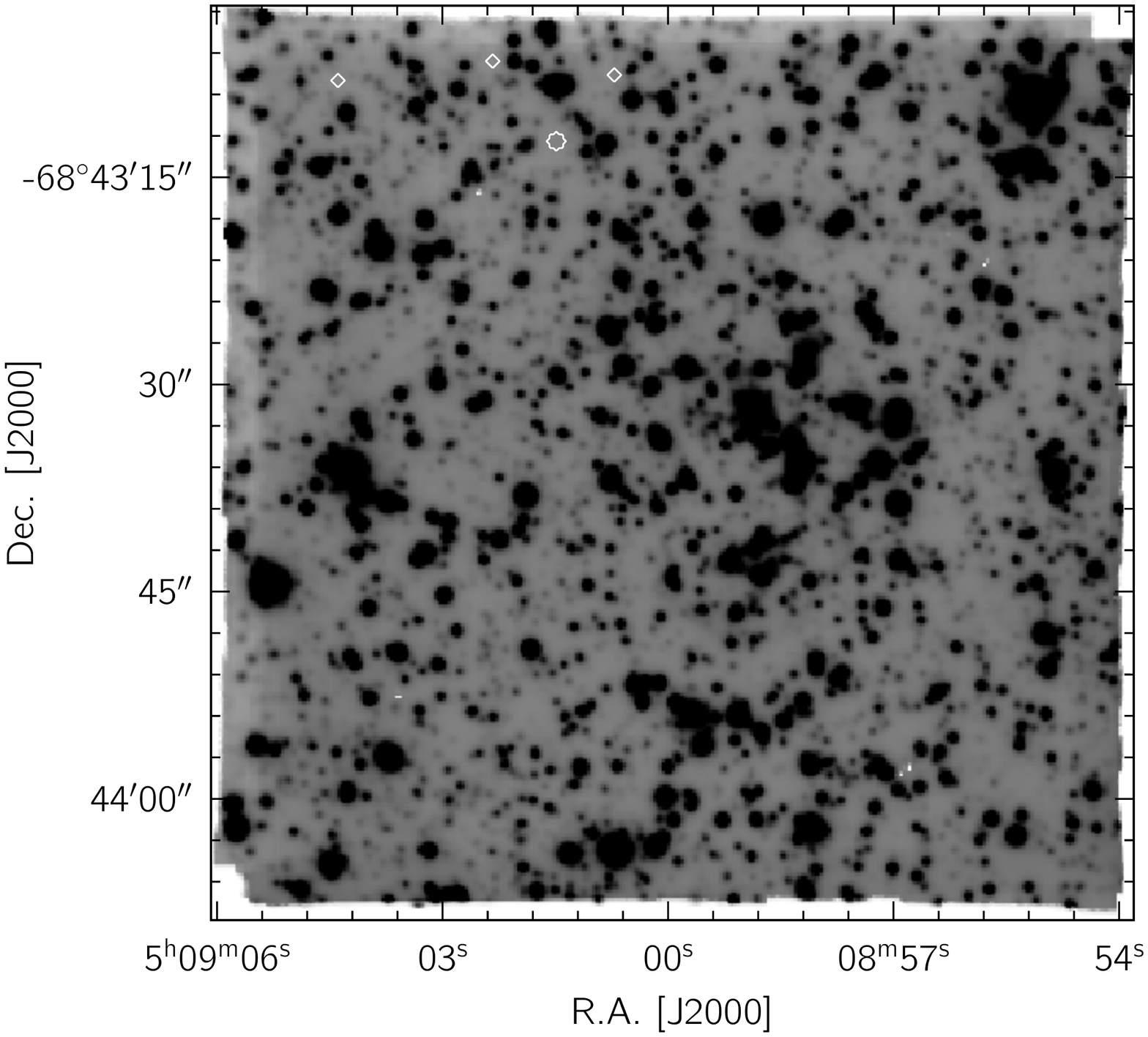}
  \vspace{-0.5cm}
  \caption{Regions used for the sky-subtraction procedure for N103B.}
  \label{sfig:f3}
\end{figure}

We correct all three combined datacubes for Galactic extinction along the line-of-sight using another dedicated brutifus routine. We assume a Fitzpatrick 1999 reddening law \cite{fitzpatrick1999a} with $R_v = 3.1$, $A_B= 0.272$ and $A_V=0.206$ (for all three systems), derived via NED from a recalibration \cite{schlafly2011a} of the infrared-based dust map of \cite{schlegel1998a}. The resulting flux correction is shown in Supplemental Fig.~7.

\begin{figure}[h!]
  \includegraphics[width=\columnwidth,trim={0 5.8cm 0 6.8cm}, clip]{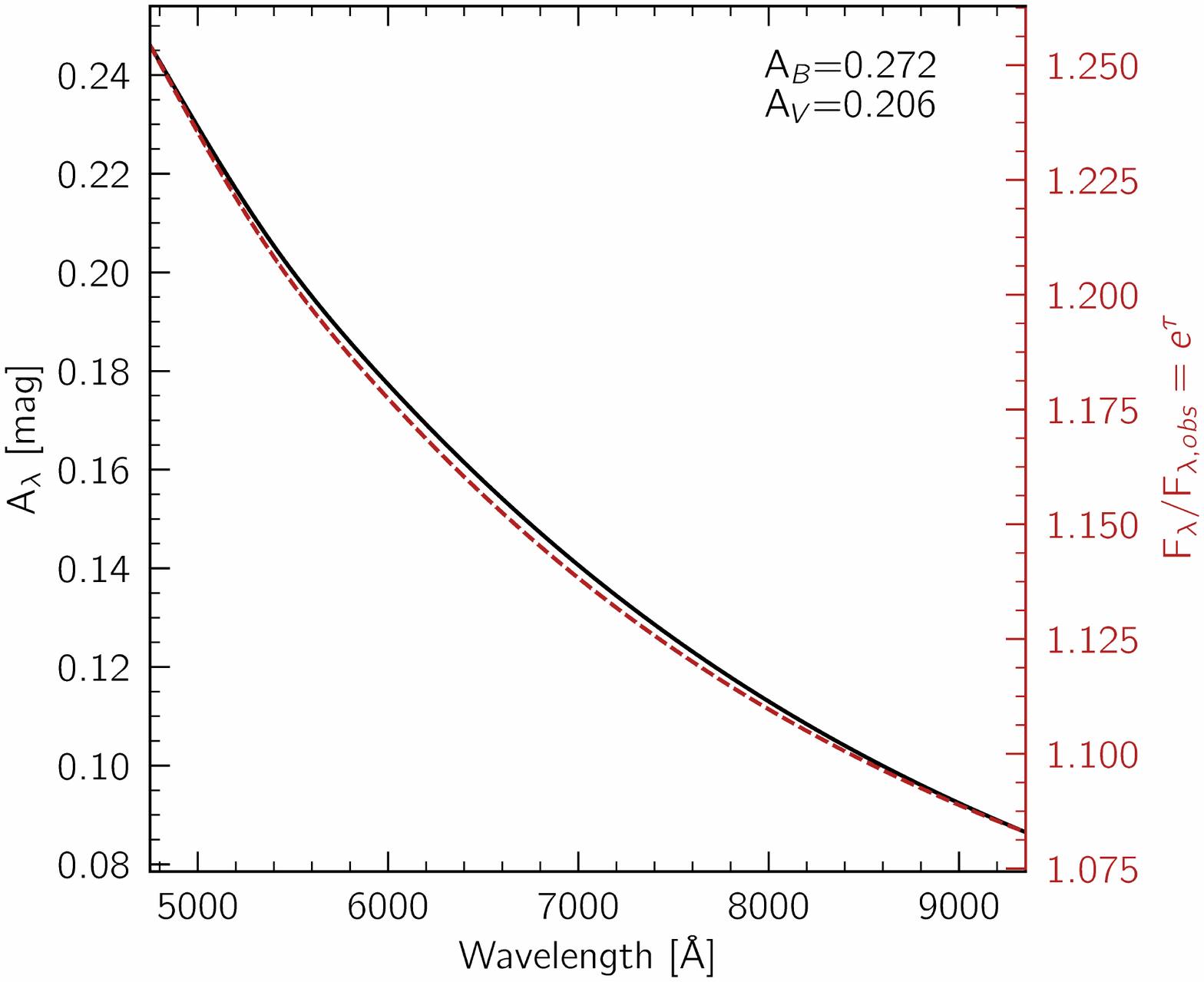}
  \caption{Wavelength dependent extinction curve applied to all three datacubes.}
  \label{sfig:f4}
\end{figure}

We also rely on brutifus to perform a crude subtraction of the stellar and nebular continuum of the three combined cubes. In practice, brutifus relies on the Locally Weighted Scatterplot Smoothing (LOWESS) algorithm \cite{cleveland1979a} to perform a non-parametric fit on a spaxel-by-spaxel basis. The advantage of this technique is that it is a) robust against the presence of emission lines, and b) can handle any type of smoothly varying continuum equally well, as illustrated by \cite{vogt2017b}.

\section{Hydrodynamics and Ionization Structure}
We model the Fe coronal line forbidden emission using the method originally pioneered by Hamilton and Sarazin \cite{hamilton1984a}. Within a framework of analytic hydrodynamics describing the SNR evolution \cite{truelove1999a}, we integrate equations for the time dependent ionization balance between the forward or reverse shocks. Our full method is described in \cite{laming2003a,hwang2012a} who coined the acronym BLASt Propagation in Highly EMitting EnviRonment (BLASPHEMER). 

Here we concentrate on Type Ia SNRs expanding into a uniform density interstellar medium (ISM). We take a core-envelope ejecta density profile, where the uniform density ejecta core is surrounded by an envelope with density proportional to r$^{-7}$. In all cases, the most highly ionized ejecta are found at the core-envelope boundary. We assume collisionless electron heating to $10^{6}$\,K ahead of the shock, following \cite{ghavamian2007a}, followed by heating by adiabatic compression and Coulomb equilibration with the ions.

0519: We assume an ejecta composition of 33\% O, 12\% Si and 55\% Fe by mass, following \cite{kosenko2010a}. Ahead of the reverse shock, O is 50\% O$^{+}$ and 50\% O$^{++}$, while Si and Fe are 25\% singly ionized, 50\% doubly ionized, and 25\% triply ionized. Supplemental Figure 8 shows the results for 0519 on the left panels. The top panel shows the predicted radial extent of the Fe$^{9+}$, Fe$^{10+}$ and Fe$^{13+}$ ions. The reverse shock is predicted to be at 2.03 pc, so it can be seen the Fe$^{9+}$ comes up first, before Fe$^{10+}$ and Fe$^{13+}$ as expected, with Fe$^{13+}$ closest to the core-envelope boundary and hence the brightest due to being in the highest density. \cite{kosenko2010a} see low and high ionization Fe ejecta in X-ray emission, which we locate in even higher density ejecta, with the high ionization Fe ejecta located close to the core-envelope boundary with ionization age $n_{e}t = 3.9 \times 10^{10}\,\mathrm{cm}^{-3}\,\mathrm{s}$. The middle panel shows the electron density profile with radius, with a strong “spike” at a radius of 2.92 pc, corresponding to the core-envelope boundary. The bottom panel shows the time after explosion of reverse shock passage for ejecta at the different radii. The radius of the contact discontinuity is overestimated. \cite{truelove1999a} give no guidance on this so this has been taken from \cite{chevalier1982a}, which is more appropriate for the earlier phases on SNR evolution.

0509: A similar ejecta composition and mass to 0519 do not ionize Fe as far as Fe$^{13+}$, and would predict significantly higher relative intensity in [Fe~\textsc{x}] and [Fe\,\textsc{xi}] than is actually present. The simplest modification is to take a smaller ejecta mass $M_{ej} = 1\,\mathrm{M}_{\odot}$ that allows the reverse shock to reach the ejecta core-envelope boundary, where the maximum ionization age occurs, earlier in the SNR evolution, allowing ionization of Fe as far as Fe$^{13+}$ and Fe$^{14+}$, with [Fe\,\textsc{xiv}] emission more intense compared to [Fe\,\textsc{x}] and [Fe\,\textsc{xi}]. The right hand panels of Supplemental Figure 5 show the same plots as previously for 0519. In the top panel, Fe$^{13+}$ and Fe$^{14+}$ are seen around the core-envelope boundary, while Fe$^{9+}$ and Fe$^{10+}$ exist at lower net regions surrounding it in the envelope and core. The density profile in the middle panel is not so strongly peaked as in 0519, suggesting stronger [Fe\,\textsc{x}] and [Fe\,\textsc{xi}] emission with respect to [Fe\,\textsc{xiv}]. For completeness, the time since explosion of reverse shock passage is similarly plotted in the bottom right.

\begin{figure*}[ht!]
  \includegraphics[width=\columnwidth,trim={2cm 3.5cm 3cm 1cm}, clip]{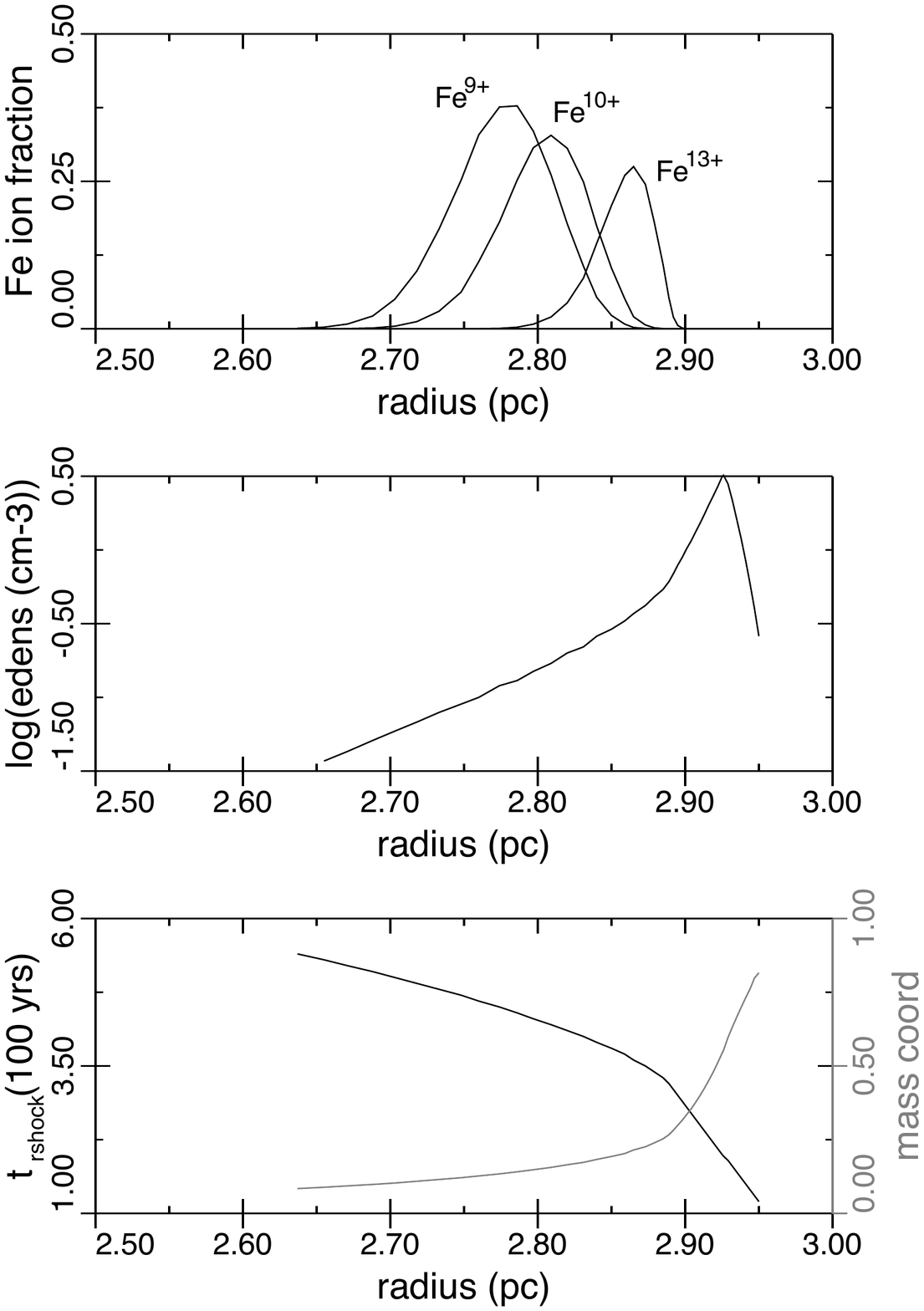}
  \includegraphics[width=\columnwidth,trim={2cm 3.5cm 3cm 1cm}, clip]{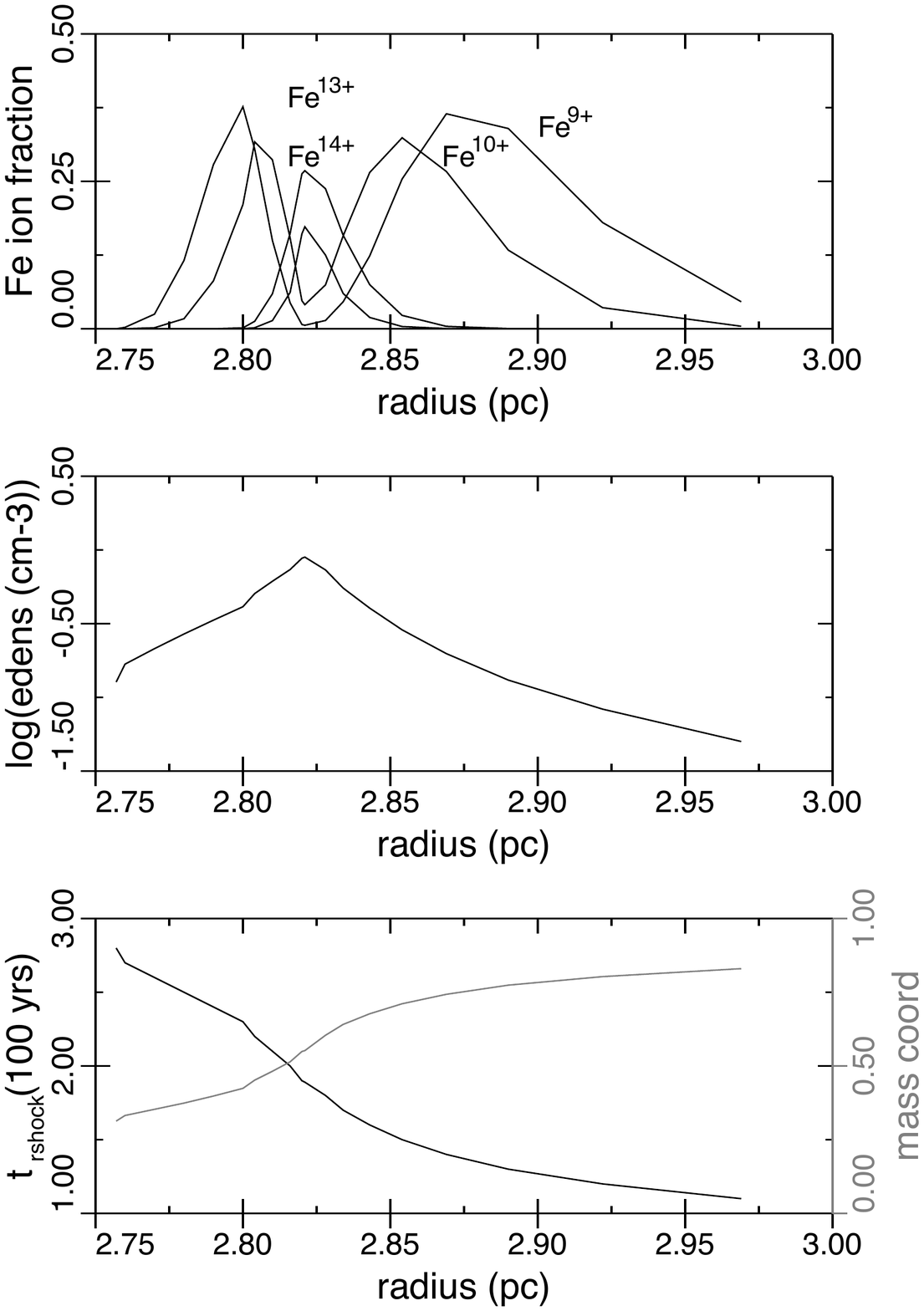}
  \caption{Predicted radial extent of the Fe$^{9+}$, Fe$^{10+}$ and Fe$^{13+}$ ions (top), electron density as a function of radius (middle), and time after explosion of reverse shock passage through the ejecta as a function of radius (bottom, left axis) and ejecta mass coordinate (bottom, right axis). Left column is for 0519 and right column is for 0509. The assumed ages are 750 years for 0519 and 310 years for 0509, respectively. }
  \label{sfig:f5}
\end{figure*}

\section{Line Emission and Fe Mass Estimates}
The Fe coronal forbidden lines are emitted from temperatures of order $ 1 - 2 \times 10^{7}\,\mathrm{K}$, well above those where they are emitted in conditions of ionization equilibrium. We take the emissivity in these lines due to electron impacts from the CHIANTI code. To this we add an emissivity due to impacts by heavy ions in the plasma, calculated using a generalization of methods in \cite{laming1996a}. The line flux, f, in photons cm$^{-2}\,\mathrm{s}^{-1}$, is given by $f = (C_{e}n_e + C_in_i) n_{Fe}V / 4 \pi d^2$, where $C_en_e + C_in_i$ is the collisional excitation rate due to scattering electrons, $n_e$, and ions $n_i$, with excitation rate coefficients $C_{e}$ and $C_i$ respectively, $n_{Fe}$ is the target ion density, V is the volume of emitting plasma, and d = 50 kpc is the distance to the SNR. The total target ion mass is 
\begin{equation}
    Vn_{Fe}m_{Fe}  = 4\pi d^2 f m_{Fe} / (C_{e}n_e + C_in_i)  
\end{equation}
                                                                              
\noindent where $m_{Fe}$ is the target ion mass. The Fe mass connected with the [Fe\,\textsc{xiv}] emission is calculated from equation 1, with $n_e$, and $n_i$ coming from the hydrodynamics and ionization evolution calculations, $C_{e}$ and $C_i$ given in Supplemental Table 2, $d = 50\,\mathrm{kpc}$ and $f$ measured as described above. Finally, we correct for the Fe ionization balance, also coming from the hydrodynamics and ionization evolution, to give a mass in all Fe charge states in the region from where [Fe\,\textsc{xiv}] photons are emitted.

In 0519 we take the low and high ionization Fe emission measures $EM_{Fe} = Vn_en_i$ quoted by \cite{kosenko2010a} and convert to masses according to $M_{Fe} = EM_{Fe} m_{Fe} / n_e$. Finally in both remnants, we assume the ejecta interior to the [Fe\,\textsc{xiv}] emission is dominated by Fe, and multiply the relevant ejecta mass coordinates (0.18 for 0519, 0.5 for 0509) by 1.4 M$_{\odot}$ (for 0519) and 1.0 M$_{\odot}$  (for 0509) to complete the mass estimate.
\vspace{-0.5cm}
\begin{table}[b]
\scriptsize
    \centering
\caption{Electron and ion impact excitation rates for [Fe\,\textsc{xiv}] 5303\AA\ and [Fe\,\textsc{x}] 6376\AA\ transitions.}
\begin{ruledtabular}
\begin{tabular}{c c c c c c}
Transition & $e^-$ 1.6e7 K & p 5e8 K & $\alpha$ 2e9 K & $C^{4+}$ 6e9 K & $O^{6+}$ 8e9 K \\
Fe\,\textsc{xiv} &$6.0\times 10^{-9}$&$1.4\times 10^{-9}$&$2.1\times 10^{-8}$&$3.1\times 10^{-8}$&$3.4\times 10^{-8}$ \\
Fe\,\textsc{x}  &$3.7\times 10^{-9}$&$1.1\times 10^{-9}$&$7.7\times 10^{-9}$&$9.7\times 10^{-9}$&$1.3\times 10^{-8}$ 
\end{tabular}
\end{ruledtabular}
\end{table}
\end{document}